\documentclass[12pt]{article}

\usepackage{amssymb}
\usepackage{amsfonts}
\usepackage{amsmath}
\usepackage{amsthm}
\usepackage[english]{babel}
\usepackage{latexsym}
\usepackage{color}
\usepackage{enumerate}
\usepackage{nicefrac}
\usepackage{mathrsfs}
\usepackage{comment}
\usepackage[hmargin=2cm,vmargin=2cm]{geometry}

\usepackage{centernot}
\usepackage{comment}
\usepackage[affil-it]{authblk}
\theoremstyle{plain}
\newtheorem{theorem}{Theorem}[section]
\newtheorem{lemma}[theorem]{Lemma}
\newtheorem{proposition}[theorem]{Proposition}
\newtheorem{corollary}[theorem]{Corollary}
\theoremstyle{definition}
\newtheorem{definition}[theorem]{Definition}
\newtheorem{remark}[theorem]{Remark}

\theoremstyle{remark}

\numberwithin{equation}{section}

\newcommand{\ba}{\begin{array}{ll}}
\newcommand{\bal}{\begin{array}{ll}}
\newcommand{\ea}{\end{array}}

\newcommand{\E}{\mathbb{E}}

\newcommand{\probp}{\mathbb{P}}

\newcommand{\R}{\mathbb{R}}
\newcommand{\N}{\mathbb{N}}
\newcommand{\Q}{\mathbb{Q}}

\newcommand{\cF}{{\mathcal{F}}}

\newcommand{\cS}{{\mathcal{S}}}
\newcommand{\cA}{\mathcal{A}}
\newcommand{\cC}{\mathcal{C}}
\newcommand{\cD}{\mathcal{D}}

\newcommand{\cP}{\mathcal{P}}

\newcommand{\cU}{\mathcal{U}}
\newcommand{\cV}{\mathcal{V}}

\newcommand{\Span}{\mathop{\rm span}\nolimits}

\newcommand{\VaR}{\mathop {\rm VaR}\nolimits}
\newcommand{\accvar}{\cA_{\VaR}}
\newcommand{\ES}{\mathop {\rm ES}\nolimits}
\newcommand{\acces}{\cA_{\ES}}

\newcommand{\CD}{\mathop {\rm DR}\nolimits}

\def\keywords{\vspace{.5em}
{\noindent\textbf{Keywords}:\,\relax%
}}

\makeatletter
\def\@fnsymbol#1{\ensuremath{\ifcase#1\or 1\or 2\or 3\or 4\or 5\or 6\or 7\or 8\else\@ctrerr\fi}}
\makeatother

\begin{document}

\title{Which eligible assets are compatible with\\ comonotonic capital requirements?}

\author{
\sc{Pablo Koch-Medina}\thanks{Email: \texttt{pablo.koch@bf.uzh.ch}}\,,
\sc{Cosimo Munari}\,\thanks{Email: \texttt{cosimo.munari@bf.uzh.ch}}
}
\affil{Center for Finance and Insurance and Swiss Finance Institute\\
University of Zurich, Switzerland}
\author{
\sc{Gregor Svindland}\,\thanks{Email: \texttt{svindla@mathematik.uni-muenchen.de}}
}
\affil{Mathematics Institute, LMU Munich, Germany}

\date{\today}

\maketitle

\begin{abstract}\noindent
Within the context of capital adequacy, we study comonotonicity of risk measures in terms of the primitives of the theory: acceptance sets and eligible, or reference, assets. We show that comonotonicity cannot be characterized by the properties of the acceptance set alone and heavily depends on the choice of the eligible asset. In fact, in many important cases, comonotonicity is only compatible with risk-free eligible assets. The incompatibility with risky eligible assets is systematic whenever the acceptability criterion is based on Value-at-Risk or any convex distortion risk measure such as Expected Shortfall. These findings qualify and arguably call for a critical appraisal of the meaning and the role of comonotonicity within a capital adequacy context.
\end{abstract}




\parindent 0em \noindent


\section{Introduction}

The theory of acceptance sets and risk measures occupies an important place in current debates about solvency regimes in both the insurance and the banking world. A variety of theoretical properties of risk measures have been studied since the seminal publication by Artzner et al.~(1999), among which the property of comonotonicity has received considerable attention. Comonotonic risk measures were first studied by Kusuoka~(2001) and Delbaen~(2002) in the mathematical finance literature and by Dhaene et al.~(2002) in the actuarial literature. We refer to F\"{o}llmer and Schied~(2011) and the references therein for a comprehensive treatment of comonotonic risk measures and to Denuit et al.~(2005) and McNeil et al.~(2015) for a thorough discussion on comonotonicity with a view towards applications in finance and insurance.

\medskip

To highlight the message of our paper, we start by recalling the concept of comonotonic risk measures in a rather informal way. Consider a one-period economy and assume that {\em capital positions} --- assets net of liabilities --- of financial institutions at the terminal date are represented by random variables belonging to a suitable ordered vector space which, in line with much of the risk measure literature, we take to be the space $L^\infty$ of essentially bounded random variables (over a given probability space). In a capital adequacy context, {\em risk measures} are interpreted as capital requirement rules and are used to determine the amount of capital a company has to hold as a regulatory buffer against unexpected future losses (we use the term ``regulatory'' in a loose sense to encompass any externally or internally imposed requirement). Mathematically, a risk measure can be represented by a map $\rho:L^\infty\to\R$. We say that a risk measure $\rho$ is {\em comonotonic} if it is additive on random variables that are {\em comonotone}, i.e.\ random variables that, being increasing functions of a common risk driver, are ``perfectly positively dependent''. The main idea behind comonotonic risk measures is that merging two comonotone positions will not result in a diversification benefit and, hence, the amount of risk capital required for the aggregated position should correspond to the sum of the individual capital requirements. This interpretation is particularly appealing if one requires $\rho$ to be {\em subadditive} so that the sum of the individual capital requirements always constitutes an upper bound for the capital requirement of a diversified position. This is the standard argument put forward in the literature to argue that comonotonicity might be a natural and desirable normative requirement in a capital adequacy framework, see e.g.\ the section on comonotonic risk measures in F\"{o}llmer and Schied~(2011).

\smallskip

The actuarial literature on capital adequacy has mainly focused on comonotonicity, especially in conjunction with subadditivity, to derive approximations and bounds for capital requirements of aggregated positions. We refer to Dhaene et al.~(2002) and to the survey article by Dhaene et al.~(2006) for a detailed presentation of such techniques. Note that, as already pointed out in Embrechts et al.~(2002), the highest capital requirement under a comonotonic risk measure for an aggregated position with given marginals need not be attained with the comonotonic copula unless the risk measure is subadditive. We refer to Embrechts et al.~(2005), Kaas et al.~(2009), and to the survey article by Embrechts et al.~(2013) for a number of results about worst bounds for Value-at-Risk.

\medskip

The objective of this paper is to investigate the property of comonotonicity for the broad class of risk measures introduced in the paper by Artzner et al.~(1999). The fundamental idea in that paper was to provide an operational definition of risk measures by emphasizing the role of two basic primitive objects:
\begin{itemize}
  \item the {\em acceptance set} $\cA\subset L^\infty$, representing the set of capital positions that are deemed acceptable from a  regulatory perspective, and
  \item the {\em eligible asset} $S=(S_0,S_1)$, representing a liquidly traded financial asset with price $S_0>0$ and payoff $S_1\in L^\infty_+$ used to reach acceptability.
\end{itemize}
The acceptance set plays the role of a {\em capital adequacy test} and is used discriminate between financial institutions that are adequately capitalized, i.e.\ whose capital position belongs to $\cA$, and those that are inadequately capitalized, i.e.\ whose capital position does not belong to $\cA$. If a company does not pass the capital adequacy test, then its management needs to implement a pre-specified {\em remedial action}, namely raising capital and investing it into the eligible asset, to become acceptable. To achieve a sufficient level of generality we will not fix a market model but view any pair $S=(S_0,S_1)$ as a potential liquidly traded asset, i.e.\ as a candidate for an eligible asset. It is worth emphasizing that the primary interest of regulators should be {\em that} the capital position of a financial institution is acceptable and, in principle, they should not be concerned about {\em how} the company reaches acceptability. In particular, as observed in Artzner et al.~(2009), it may be less costly for the company to reach acceptability by means of a risky asset instead of a risk-free one, which potentially may not even exist.

\smallskip

The {\em risk measure associated with $\cA$ and $S$} is defined for every $X\in L^\infty$ by setting
\[
\rho_{\cA,S}(X):=\inf\{m\in\R \mid \, X+\tfrac{m}{S_0}S_1\in\cA\}.
\]
If $S$ is the {\em cash asset}, i.e.\ $S_0=S_1=1$, we simply write
\[
\rho_\cA(X):=\inf\{m\in\R \mid\, X+m\in\cA\}.
\]
When finite, the quantity $\rho_{\cA,S}(X)$ has a very clear operational interpretation. If positive, $\rho_{\cA,S}(X)$  is the ``minimum'' amount of capital that, if raised and invested in the eligible asset, makes the position $X$ acceptable. If negative, $-\rho_{\cA,S}(X)$ is the maximum amount of capital that can be returned to the owners  while retaining acceptability. As discussed in Section~\ref{two}, in this paper we rule out the situations where $\rho_{\cA,S}$ is not finitely valued by making suitable assumptions on the payoff $S_1$.

\medskip

Our aim is to characterize comonotonicity for risk measures of the form $\rho_{\cA,S}$. This problem has so far been addressed in the literature only for the special case of a {\em risk-free} eligible asset. As established in Proposition~\ref{thm:essrhoA}, a risk measure of the form $\rho_{\cA,S}$ is comonotonic if and only if the corresponding cash-based risk measure $\rho_{\cA}$ is comonotonic and coincides with $\rho_{\cA,S}$ (up to a multiple). Hence, our problem boils down to investigating when a comonotonic cash-based risk measure can be represented as $\rho_{\cA,S}$ for a suitable risky eligible asset $S$. As illustrated by our examples, the comonotonicity of $\rho_\cA$ is not sufficient for $\rho_{\cA,S}$ to be comonotonic. Consequently, the question is not settled by the existing literature.

\medskip

It is well-known, see e.g.\ the remark after Definition~2.1 in Delbaen~(2002) or Remark~2.2 in F\"{o}llmer and Schied~(2002), that risk measures of the form $\rho_{\cA,S}$ can be expressed in terms of cash-based risk measures applied to ``discounted'' positions. Indeed, if we use $S$ as a new num\'{e}raire, one can write
\[
\rho_{\cA,S}(X) = S_0\rho_{\cA'}(X')
\]
for every $X\in L^\infty$,  where $X'=X/S_1$ is the ``discounted'' version of $X$ and $\cA'=\{X/S_1\mid X\in\cA\}$. The new risk measure $\rho_{\cA'}$ is still defined on $L^\infty$ provided that $S_1$ is bounded away from zero. As a result, it is natural to ask whether the study of comonotonicity for $\rho_{\cA,S}$ cannot be, after all, reduced to the standard cash-based case. However, comonotonicity is not preserved by a change of num\'{e}raire so that the comonotonicity of $\rho_{\cA,S}$ (in the ``undiscounted'' world) neither implies nor is implied by that of $\rho_{\cA'}$ (in the ``discounted'' world). Hence, discounting does not help to address our problem.

\medskip

The main result on comonotonicity is recorded in Theorem~\ref{thm: characterization comonotonicity risk measures}. This result shows that comonotonicity is not per se incompatible with a risky eligible asset. However, as illustrated by Corollary~\ref{cor: characterization comonotonicity convex risk measures} and Corollary~\ref{cor: comonotonicity under pointedness}, the class of eligible assets under which $\rho_{\cA,S}$ is comonotonic is quite restrictive for most of the common examples. For instance, whenever the acceptance set is coherent and law-invariant, it is necessary that the eligible asset is risk-free, except in the trivial case when the acceptance set consists of all those positions that have nonnegative expectation (in which case, $\rho_{\cA,S}$ is linear and, hence, comonotonic regardless of the choice of $S$). In particular, comonotonicity is never compatible with risky eligible assets under an Expected Shortfall regime. The conclusion is similar for the most prominent noncoherent case, namely the case of acceptance sets based on Value-at-Risk. Here, if the underlying probability space is nonatomic, $\rho_{\cA,S}$ cannot be comonotonic for any risky eligible asset. In the case that the underlying probability space is finite, comonotonicity is only compatible with eligible assets that are close to risk-free as is made precise in Lemma~\ref{prop:var comonotonic nonconstant S}.

\medskip

The structure of the paper is as follows. In Section~\ref{two} we describe the underlying framework and establish our main results. In Section~\ref{three} these results are applied to a variety of examples, which include the most commonly used acceptability criteria. Section~\ref{four} concludes.


\section{Risk measures and comonotonicity}
\label{two}

In this section we provide a comprehensive study of comonotonicity for the class of risk measures introduced by Artzner et al.~(1999).

\subsubsection*{The mathematical setup}

Throughout the whole paper we consider a fixed probability space $(\Omega,\cF,\probp)$. All probabilistic statements such as ``almost sure (a.s)'', ``essential'', and the like are silently understood to be with respect to $\probp$. We always equip $\R$ with the Borel $\sigma$-algebra. Recall that a random variable $X:\Omega\to\R$ is said to be \textit{essentially bounded} whenever
\[
\|X\|:=\inf\{m\in\R\mid \probp(|X|>m)=0\}<\infty.
\]
The constant random variable with value $\lambda\in\R$, which is still denoted by $\lambda$, and the indicator function of an event $E\in\cF$, denoted by $1_E$, are examples of essentially-bounded random variables. The vector space $L^{\infty}:=L^{\infty}(\Omega,\cF,\probp)$ consists of the equivalence classes of essentially-bounded random variables with respect to a.s.\ equality. As is customary, we identify equivalence classes in $L^\infty$ with any of their representatives. The space $L^\infty$ is a Banach lattice when equipped with the above norm and with the partial order
\[
X\geq Y \ :\iff \ \probp(X\geq Y)=1.
\]
Topological concepts for subsets $L^\infty$ are always understood to be with respect to the above norm. For a sequence $(X_n)\subset L^\infty$ converging to $X\in L^\infty$ we write $X_n\to X$. The set of positive elements of $L^\infty$ is given by
\[
L^\infty_+:= \{X\in L^\infty\mid X\ge 0\}.
\]
Moreover, for every $X\in L^\infty$ we denote by $\probp_X$ the probability law of $X$ under $\probp$.

\smallskip

The next two definitions collect for easy reference a variety of properties of functionals and sets of random variables that will be freely used throughout the paper.
\begin{definition}
A functional $\rho:L^\infty\to\R$ may satisfy the following properties:
\begin{enumerate}[(1)]
  \item $\rho(\lambda X+(1-\lambda)Y)\leq\lambda\rho(X)+(1-\lambda)\rho(Y)$ for all $\lambda\in[0,1]$ and $X,Y\in L^\infty$ ({\em convexity}).
  \item $\rho(\lambda X)=\lambda\rho(X)$ for all $\lambda\in\R_+$ and $X\in L^\infty$ ({\em positive homogeneity}).
  \item $\rho(X+Y)\leq\rho(X)+\rho(Y)$ for all $X,Y\in L^\infty$ ({\em subadditivity}).
  \item $\rho(X)\leq\rho(Y)$ for all $X,Y\in L^\infty$ with $X\geq Y$ ({\em decreasing monotonicity}).
  \item $\rho(X)=\rho(Y)$ for all $X,Y\in L^\infty$ such that $\probp_X=\probp_Y$ ({\em law-invariance}).
  \item $|\rho(X)-\rho(Y)|\leq c\|X-Y\|$ for all $X,Y\in L^\infty$ and some $c\in[0,\infty)$ ({\em Lipschitz continuity}).
\end{enumerate}
\end{definition}

\medskip

\begin{definition}
A set $\cA\subset L^\infty$ may satisfy the following properties:
\begin{enumerate}[(1)]
  \item $\lambda\cA+(1-\lambda)\cA\subset\cA$ for all $\lambda\in[0,1]$ ({\em convexity}).
  \item $\lambda \cA\subset\cA$ for all $\lambda\in[0,\infty)$ ({\em conicity}).
  \item $\cA+\cA\subset\cA$ ({\em closedness under addition}).
  \item $\cA\cap(-\cA)\subset\{0\}$ ({\em pointedness}).
  \item $\cA+L^\infty_+\subset\cA$ ({\em increasing monotonicity}).
  \item $X\in\cA$ for all $X\in L^\infty$ such that $\probp_X=\probp_Y$ for some $Y\in\cA$ ({\em law-invariance}).
\end{enumerate}
\end{definition}

\smallskip

\begin{remark}
Recall that a positively  homogeneous functional $\rho:L^\infty\to\R$ is convex if, and only if, it is subadditive. Similarly, a cone is convex if, and only if, it is closed under addition.
\end{remark}

\smallskip

We conclude this introductory section by recalling the notion of comonotonicity. The terminology was introduced in Schmeidler~(1986) and the characterization in terms of a common risk driver can be found in Denneberg~(1994). We refer to the introduction for a financial interpretation in the context of capital adequacy.

\begin{definition}
We say that $X,Y\in L^\infty$ are {\em comonotone} whenever there is a $\probp\otimes\probp$-null set $N\subset\cF\otimes\cF$ such that
\[
(X(\omega)-X(\omega'))(Y(\omega)-Y(\omega'))\ge 0\quad\mbox{for all} \ (\omega,\omega')\in\Omega\times \Omega \setminus N
\]
This is equivalent to the existence of $Z\in L^\infty$ and of two increasing functions $f,g:\R\to\R$ satisfying
\[
X=f(Z) \ \ \ \mbox{and} \ \ \ Y=g(Z).
\]
A set $\cC\subset L^\infty$ is said to be {\em comonotonic} if $X$ and $Y$ are comonotone for any choice of $X,Y\in\cC$. A functional $\rho:L^\infty\to\R$ is called {\em comonotonic} whenever
\[
\rho(X+Y) = \rho(X)+\rho(Y)
\]
for all comonotone $X,Y\in L^\infty$.
\end{definition}

\medskip

The following result collects some important properties of comonotonic maps.

\begin{proposition}
\label{prop: properties comonotonicity}
Let $\rho:L^\infty\to\R$ be a nonzero comonotonic decreasing map. Then, $\rho(1)<0$ and the following statements hold:
\begin{enumerate}[(i)]
\item $\rho(X+m)=\rho(X)+m\rho(1)$ for every $X\in L^\infty$ and every $m\in\R$.
\item $\rho$ is Lipschitz continuous with constant $-\rho(1)$.
\item $\rho$ is positively homogeneous.
\end{enumerate}
\end{proposition}
\begin{proof}
We easily see that comonotonicity implies $\rho(0)=0$ and, hence, $\rho(-1)=-\rho(1)$. Moreover, we have $\rho(1)\leq0$ by monotonicity. We claim that $\rho(1)<0$ holds. To prove this, assume that $\rho(1)=0$ and take $X\in L^\infty$. Since $1$ is comonotone with itself and $a\geq X\geq-a$ for $a\in\N$ large enough, we obtain
\[
0 = a\rho(1) = \rho(a)\leq \rho(X) \leq  \rho(-a) = a\rho(-1) = 0
\]
which is only possible if $\rho$ is the zero function. In conclusion, we see that $\rho(1)<0$ must hold.

\smallskip

Now, take $X\in L^\infty$ and two integers $a,b\in\N$. Since every random variable is comonotone with itself we have
\[
\rho(\tfrac{a}{b}X) =
a\rho(\tfrac{1}{b}X) =
\tfrac{a}{b}\rho(b\tfrac{1}{b}X) =
\tfrac{a}{b}\rho(X).
\]
As a result, for every rational number $m\in\Q$ it follows that
\[
\rho(X+m) = \rho(X)+\rho(m) = \rho(X)-m\rho(-1),
\]
where we used that $m$ is comonotone with $X$ and that $\rho(-1)=-\rho(1)$.

\smallskip

Take now an arbitrary $m\in\R$ and let $(m^+_n)$ and $(m_n^-)$ be sequences of rational numbers converging to $m$ from above and below, respectively. Using the monotonicity of $\rho$ we finally obtain
\[
\rho(X)-m_n^+\rho(-1) = \rho(X+m_n^+) \leq \rho(X+m) \leq \rho(X+m_n^-) = \rho(X)-m_n^-\rho(-1)
\]
for any $n\in\N$. This implies that $\rho(X+m)=\rho(X)-m\rho(-1)$ and establishes~{\em (i)}.

\smallskip

We have already established positive homogeneity for positive rational numbers. Take now $\lambda\in(0,\infty)$ and let $(\lambda^+_n)$ and $(\lambda_n^-)$ be sequences of positive rational numbers converging to $\lambda$ from above and below, respectively. In view of~{\em (i)} we can assume that $X\in L^\infty_+$ without loss of generality. Then, the monotonicity of $\rho$ yields
\[
\lambda^+_n\rho(X) = \rho(\lambda^+_nX) \leq \rho(\lambda X) \leq \rho(\lambda^-_nX) = \lambda^-_n\rho(X)
\]
for every $n\in\N$. This yields $\rho(\lambda X)=\lambda\rho(X)$ and concludes the proof of~{\em (iii)}.

\smallskip

To prove Lipschitz continuity we follow the proof of Lemma~4.3 in~\cite{FoellmerSchied2011}. To that effect, take $X,Y\in L^\infty$ and note that $Y-\|X-Y\|\leq X\leq Y+\|X-Y\|$. Then, it follows immediately from monotonicity and from~{\em (i)} that
\[
\rho(Y)+\|X-Y\|\rho(-1) \geq \rho(X) \geq \rho(Y)-\|X-Y\|\rho(-1).
\]
This establishes~{\em (ii)} and concludes the proof of the proposition.
\end{proof}

\smallskip

\begin{remark}
An alternative proof of the above proposition relies on the Choquet representation in Schmeidler~(1986) and exploits the properties of Choquet integrals, see e.g.\ Proposition~5.1 in Denneberg~(1994). The above proof has the advantage of being direct.
\end{remark}

\medskip

We conclude by showing that a decreasing map is comonotonic if, and only if, it coincides with a negative linear functional on each maximal comonotonic set.

\begin{definition}
We say that $\cC\subset L^\infty$ is a {\em maximal comonotonic set} if $\cC$ is comonotonic and for every comonotonic set $\cC'\subset L^\infty$ satisfying $\cC\subset\cC'$ we have $\cC=\cC'$. The class of maximal comonotonic set is denoted by $\cC_{max}$.
\end{definition}

\smallskip

The next lemma shows that every random variable in $L^\infty$ is contained in some maximal comonotonic set. Clearly, this set is not unique, in general, since the constant random variables are contained in every comonotonic set.

\begin{lemma}
For every $X\in L^\infty$ there exists a maximal comonotonic set $\cC\in\cC_{max}$ such that $X\in\cC$.
\end{lemma}
\begin{proof}
Denote by $\cC_X$ the class of comonotonic subsets of $L^\infty$ that contain $X$. Since the set $\{X\}$ is comonotonic, this class is nonempty and can be partially ordered by inclusion. If $\cS$ is a chain in $\cC_X$, i.e.\ a subset of $\cC_X$ such that, for every $\cC_1,\cC_2\in\cS$, we either have $\cC_1\subset\cC_2$ or $\cC_2\subset\cC_1$, then it is easy to see that $\cC_{\cup}=\bigcup_{\cC\in\cS}\cC$ is a comonotonic set containing $X$. In other words, $\cC_{\cup}$ is an upper bound for $\cS$. It follows from Zorn's Lemma, see Lemma 1.7 in Aliprantis and Border~(2009), that $\cC_X$ admits a maximal element.
\end{proof}

\smallskip

\begin{proposition}
\label{prop: characterization comonotonicity decreasing maps}
For every nonzero decreasing map $\rho: L^\infty\to\R$ the following statements are equivalent:
\begin{enumerate}[(a)]
  \item For each $\cC\in\cC_{max}$ there is a positive linear functional $\pi_\cC:L^\infty\to\R$ such that $\rho=-\pi_\cC$ on $\cC$.
  \item $\rho$ is comonotonic.
\end{enumerate}
If $\rho$ is convex, we may add the following equivalent condition:
\begin{enumerate}[(c)]
  \item For each $\cC\in\cC_{max}$ there is a positive linear functional $\pi_\cC:L^\infty\to\R$ such that $\rho=-\pi_\cC$ on $\cC$ and $\rho\geq-\pi_\cC$ on $L^\infty$.
\end{enumerate}
\end{proposition}
\begin{proof}
To establish the first equivalence we only need to prove that {\em (b)} implies {\em (a)}. To this end, assume that $\rho$ is comonotonic and for every fixed $\cC\in\cC_{max}$ define $\cV=\cC-\cC$. Since $\cC$ is easily seen to be a convex cone, the set $\cV$ is a linear subspace of $L^\infty$. Now, define a map $\pi_\cC:\cV\to\R$ by setting
\begin{equation}
\label{eq:g4}
\pi_\cC(X)=-\rho(X_1)+\rho(X_2),
\end{equation}
where $X=X_1-X_2$ with $X_1,X_2\in\cC$. Note that $\pi_\cC$ is well-defined on $\cV$. Indeed, assume $X=X_1-X_2=X'_1-X'_2$ for $X_1,X_2,X'_1,X'_2\in\cC$. Since $\cC$ is a convex cone, the random variable $X_1+X'_2=X'_1+X_2$ belongs to $\cC$ and therefore
\[
\rho(X_1)+\rho(X'_2) = \rho(X_1+X'_2) = \rho(X'_1+X_2) = \rho(X'_1)+\rho(X_2)
\]
by comonotonicity. Moreover, $\pi_\cC$ is clearly linear and it is also positive on $\cV$. The latter is due to the fact that $X=X_1-X_2\in \cV\cap L^\infty_+$ implies $X_1\geq X_2$ and thus $\rho(X_1)\leq \rho(X_2)$ by the monotonicity of $\rho$, which finally implies that $\pi_\cC(X)=-\rho(X_1)+\rho(X_2)\geq 0$. Since $1\in\cV$, we can apply Kantorovich's extension criterion, see Theorem~8.32 in Aliprantis and Border~(2006), to conclude that there exists a (not necessarily unique) positive linear extension of $\pi_\cC$ to $L^\infty$.

\smallskip

To conclude the proof, assume that $\rho$ is convex and comonotonic and let $\cC$ be a fixed maximal comonotonic set in $L^\infty$. Then, for every $X=X_1-X_2$ with $X_1,X_2\in\cC$ we have
\[
\pi_\cC(X)+\pi_\cC(X_2) = \pi_\cC(X_1) = -\rho(X_1) \geq -\rho(X)-\rho(X_2) = -\rho(X)+\pi_\cC(X_2),
\]
where $\pi_\cC$ is the functional defined in~\eqref{eq:g4} and the inequality follows from the subadditivity of $\rho$ (which is positively homogeneous by Proposition~\ref{prop: properties comonotonicity}). This implies that $\rho\geq-\pi_\cC$ on the linear subspace $\cV$. Now, the Hahn-Banach extension theorem, see Theorem~5.53 in Aliprantis and Border~(2006), ensures the existence of a linear extension of $\pi_\cC$ to the whole of $L^\infty$, which we also denote by $\pi_\cC$, such that $\rho\geq-\pi_\cC$ on the entire $L^\infty$. In particular, $\pi_\cC$ is positive because for all $X\in L^\infty_+$ we have $\pi_\cC(X)\geq-\rho(X)\geq 0$ as $\rho$ is decreasing with $\rho(0)=0$. This implies that {\em (b)} implies {\em (c)} and concludes the proof.
\end{proof}

\smallskip

\begin{remark}
\label{rem: comonotonic sets and permutations}
(i) Note that, being positive, every linear functional $\pi_\cC$ is automatically continuous by Theorem~9.6 in Aliprantis and Border~(2006).

\smallskip

(ii) Let $ \Omega=\{\omega_1,\dots,\omega_N\}$ and assume that $\cF$ is the discrete $\sigma$-algebra and $\probp(\omega_i)>0$ for every $i\in\{1,\dots,N\}$. In this setting, the maximal comonotonic sets have the form
\[
\cC_p = \{X\in L^\infty\mid X(\omega_{p(k)})\leq X(\omega_{p(k+1)}), \ k\in\{1,\dots,N-1\}\}
\]
for some bijection $p:\{1,\dots,N\}\to\{1,\dots,N\}$. In particular, we have precisely $N!$ maximal comonotonic sets. To see this, note that every set of the form $\cC_p$ is indeed a maximal comonotonic set. Conversely, let $\cC$ be a maximal comonotonic set in $L^\infty$ and note that $\cC$ must contain a random variable $Z\in L^\infty$ such that $Z(\omega_i)\neq Z(\omega_j)$ for all $i,j\in\{1,\dots,N\}$ with $i\neq j$. Let $p$ be a bijection of $\{1,\dots,N\}$ which sorts the values of $Z$ in such a way that $Z(\omega_{p(k)})<Z(\omega_{p(k+1)})$ for all $k\in\{1,\dots,N-1\}$. Then, every $X\in\cC$ must necessarily satisfy $X(\omega_{p(k)})\leq X(\omega_{p(k+1)})$ for $k\in\{1,\dots,N-1\}$ so that $\cC\subset\cC_p$. Hence, by maximality, we have $\cC=\cC_p$.

\smallskip

(iii) It is clear that a similar characterization of maximal comonotonic sets as in (ii) also holds if $\Omega$ is a countable set.
\end{remark}

\subsubsection*{Introducing risk measures}

We consider a one-period economy with dates $t=0$ and $t=1$ in which future uncertainty is modelled by the probability space $(\Omega,\cF,\probp)$. The capital of a financial institution at time $1$, i.e.\ the value of the company's assets net of liabilities, is represented by a random variable in $L^\infty$. Every element of $L^\infty$ will be referred to as a {\em capital position}.

\begin{definition}
\begin{enumerate}[(1)]
  \item Any nonempty proper subset $\cA\subset L^\infty$ that is closed and increasing is said to be an {\em acceptance set}.
  \item Any couple $S=(S_0,S_1)\in(0,\infty)\times L^\infty_+$ with $S_1\geq\varepsilon$ a.s. for some $\varepsilon>0$ is called an {\em eligible asset}. We say that $S$ is {\em risk-free} whenever $S_1$ is constant. Otherwise, we say that $S$ is {\em risky}.
  \item The {\em risk measure} associated to $\cA$ and $S$ is the map $\rho_{\cA,S}:L^\infty\to\R$ defined for every $X\in L^\infty$ by
\[
\rho_{\cA,S}(X) := \inf\{m\in\R\mid X+\tfrac{m}{S_0}S_1\in\cA\}.
\]
(The requirement that $S_1$ be bounded away from zero ensures, by Farkas et al.~(2014a), that $\rho_{\cA,S}$ is finitely valued). If $S=(1,1)$, then we simply write
\[
\rho_\cA(X) := \inf\{m\in\R\mid X+m\in\cA\}.
\]
\end{enumerate}
Throughout the remainder of the paper we fix an acceptance set $\cA$ and an eligible asset $S$.
\end{definition}

\smallskip

\begin{definition}
\begin{enumerate}[(1)]
  \item We say that a map $\rho:L^\infty\to\R$ is {\em $S$-additive} whenever
\[
\rho_{\cA,S}(X+\lambda S_1)=\rho_{\cA,S}(X)-\lambda S_0
\]
for all $X\in L^\infty$ and $\lambda\in\R$. If $S=(1,1)$, then we speak of {\em cash-additivity}.
  \item For every $\rho:L^\infty\to\R$ we set $\cA(\rho):=\{X\in L^\infty\mid  \rho(X)\le 0\}$.
\end{enumerate}
\end{definition}

\medskip

As a result of the monotonicity of the acceptance set and the linearity of the pricing rule, the risk measure $\rho_{\cA,S}$ is easily seen to enjoy the following fundamental properties, which will be freely used in the sequel; see Artzner et al.~(1999) and, for the present setting, Farkas et al.~(2014a).

\medskip

\begin{proposition}
\label{prop: usual properties}
The risk measure $\rho_{\cA,S}$ satisfies the following properties:
\begin{enumerate}[(i)]
  \item $\rho_{\cA,S}$ is $S$-additive.
  \item $\rho_{\cA,S}$ is decreasing.
  \item $\cA(\rho_{\cA,S})=\cA$.
\end{enumerate}
\end{proposition}

\subsubsection*{Characterizing comonotonicity}

We start our investigation of comonotonic risk measures of the form $\rho_{\cA,S}$ by highlighting that {\em any} comonotonic decreasing functional can be expressed as the risk measure associated to a conic acceptance set and to a risk-free asset.

\begin{lemma}
\label{prop: comonotonicity implies risk measure}
Let $\rho:L^\infty\to\R$ be a nonzero comonotonic decreasing map. Then, $\rho(1)<0$ and
\[
\rho(X) = \rho_{\cA(\rho),R}(X)
\]
for every $X\in L^\infty$, where $\cA(\rho)$ is a conic acceptance set and $R=(-\rho(1),1)$.
\end{lemma}
\begin{proof}
It follows from Proposition~\ref{prop: properties comonotonicity} that $\rho(1)<0$ holds and $\rho$ is $R$-additive. Now, take an arbitrary $X\in L^\infty$. Since the condition $X+\tfrac{m}{R_0}R_1\in\cA(\rho)$ is clearly equivalent to $\rho(X)\leq m$ for every $m\in\R$ by $R$-additivity, we easily see that
\[
\rho(X) = \inf\{m\in\R\mid X+\tfrac{m}{R_0}R_1\in\cA(\rho)\} = \rho_{\cA(\rho),R}(X).
\]
That $\cA(\rho)$ is a conic acceptance set is a direct consequence of Proposition~\ref{prop: properties comonotonicity}.
\end{proof}

\medskip

This result allows us to provide necessary and sufficient conditions for a risk measure $\rho_{\cA,S}$ to be comonotonic in terms of properties of the acceptance set and of the eligible asset. First of all, we show that $\rho_{\cA,S}$ cannot be comonotonic unless the cash-additive risk measure $\rho_\cA$ is itself comonotonic.

\begin{proposition}
\label{thm:essrhoA}
Assume that $\rho_{\cA,S}$ is comonotonic. Then, $\rho_\cA$ is comonotonic and
\[
\rho_{\cA,S}(X) = \rho_{\cA,R}(X) = -\rho_{\cA,S}(1)\rho_\cA(X)
\]
for every $X\in L^\infty$, where $R=(-\rho_{\cA,S}(1),1)$.
\end{proposition}
\begin{proof}
Recall from Proposition~\ref{prop: usual properties} that $\cA(\rho_{\cA,S})=\cA$. Hence, Lemma~\ref{prop: comonotonicity implies risk measure} implies that
\[
\rho_{\cA,S}(X) = \rho_{\cA,R}(X) = \inf\{m\in\R\mid X-\tfrac{m}{\rho_{\cA,S}(1)}\in\cA\} = -\rho_{\cA,S}(1)\rho_\cA(X)
\]
for all $X\in L^\infty$. This also shows that $\rho_\cA$ is comonotonic.
\end{proof}

\medskip

It follows from the above result that a necessary condition for the comonotonicity of $\rho_{\cA,S}$ is that, besides being additive with respect to the eligible asset $S$, the risk measure $\rho_{\cA,S}$ also needs to be additive with respect to the risk-free asset $R=(-\rho_{\cA,S}(1),1)$ (recall that, despite our terminology, $R$ is not necessarily traded in the market). This naturally leads to the study of the equality of two risk measures $\rho_{\cA,S}$ and $\rho_{\cA,R}$ with respect to the same acceptance set but different eligible assets. The following lemma establishes a necessary and sufficient condition for equality. Here, we denote by $\Span(X)$ the linear space generated by $X\in L^\infty$.

\begin{lemma}
\label{lem: equality risk measures}
For every eligible asset $R$ the following statements are equivalent:
\begin{enumerate}[(a)]
  \item $\rho_{\cA,S}(X)=\rho_{\cA,R}(X)$ for every $X\in L^\infty$.
  \item $\cA+\Span\big(\tfrac{S_1}{S_0}-\tfrac{R_1}{R_0}\big)=\cA$.
\end{enumerate}
\end{lemma}
\begin{proof}
To prove that {\em (a)} implies {\em (b)}, assume that $\rho_{\cA,S}=\rho_{\cA,R}$. Clearly, we only need to prove the inclusion ``$\subset$'' in {\em (b)}. To this effect, take any $X\in L^\infty$ and $m\in\R$ and note that
\[
\rho_{\cA,S}\big(X+m\big(\tfrac{S_1}{S_0}-\tfrac{R_1}{R_0}\big)\big) = \rho_{\cA,S}(X),
\]
where we used that $\rho_{\cA,S}$ is additive with respect to both $S$ and $R$ by assumption. The assertion now follows from $\cA(\rho_{\cA,S})=\cA$.

\smallskip

To prove the converse implication, take $X\in L^\infty$ and assume that {\em (b)} holds. If $X+\tfrac{m}{S_0}S_1\in\cA$ for some $m\in\R$, then we easily see that
\[
X+\tfrac{m}{R_0}R_1 = X+\tfrac{m}{S_0}S_1-m\big(\tfrac{S_1}{S_0}-\tfrac{R_1}{R_0}\big) \in \cA
\]
as well. This implies that $\rho_{\cA,R}(X)\leq\rho_{\cA,S}(X)$. Since the argument is symmetric in $S$ and $R$, we conclude that $\rho_{\cA,S}=\rho_{\cA,R}$.
\end{proof}

\smallskip

\begin{remark}
The above lemma could be derived from Proposition~5.1 in Farkas et al.~(2014a), see also Proposition~1-a in Artzner et al.~(2009) in a convex setting. We have provided a short proof for the sake of completeness.
\end{remark}

\medskip

We are now in a position to state our first characterization of comonotonicity for risk measures of the form $\rho_{\cA,S}$. This result identifies which combinations of acceptance sets and eligible assets give rise to comonotonicity. In particular, it shows that comonotonicity cannot be characterized by the properties of the acceptance set alone: For a given acceptance set, a risk measure $\rho_{\cA,S}$ is comonotonic only for special choices of the eligible asset $S$.

\begin{theorem}
\label{thm: characterization comonotonicity risk measures}
Assume that $\rho_\cA$ is comonotonic. Then, the following statements are equivalent:
\begin{enumerate}[(a)]
  \item $\rho_{\cA,S}$ is comonotonic.
  \item $\cA\pm\big(1+\tfrac{\rho_{\cA,S}(1)}{S_0}S_1\big)\subset\cA$.
\end{enumerate}
In particular, $\rho_{\cA,S}$ is comonotonic only if $S_1+\tfrac{S_0}{\rho_{\cA,S}(1)}\in\cA\cap(-\cA)$.
\end{theorem}
\begin{proof}
First, assume that {\em (a)} holds so that $\rho_{\cA,S}$ is comonotonic. Then, by Proposition~\ref{thm:essrhoA}, we have
\[
\rho_{\cA,S}(X)=-\rho_{\cA,S}(1)\rho_\cA(X) = \rho_{\cA,R}(X)
\]
for every $X\in L^\infty$, where $R=(-\rho_{\cA,S}(1),1)$. Hence, Lemma~\ref{lem: equality risk measures} immediately yields {\em (b)}.

\smallskip

Conversely, assume that {\em (b)} holds. As $\rho_\cA$ is comonotonic, $\cA$ must be conic by Lemma~\ref{prop: comonotonicity implies risk measure}. Moreover, note that $\rho_{\cA,S}(1)<0$ by virtue of the same lemma. Since condition {\em (b)} is then easily seen to be equivalent to
\[
\cA+\Span\big(\tfrac{S_1}{S_0}-\tfrac{1}{-\rho_{\cA,S}(1)}\big) \subset \cA,
\]
we can rely on Lemma~\ref{lem: equality risk measures} to conclude that $\rho_{\cA,S}$ coincides, up to the constant $-\rho_{\cA,S}(1)$, with the comonotonic risk measure $\rho_\cA$ and is therefore itself comonotonic. This establishes {\em (a)} and concludes the proof of the equivalence. The last assertion follows immediately because $0\in\cA$ by conicity.
\end{proof}

\medskip

To interpret condition {\em (b)} in the above theorem assume that $\rho_{\cA,S}(1)=-1$ and that cash is a traded asset (i.e.\ a risk-free asset with zero interest rate). In this case, the condition boils down to
\[
\cA\pm\big(1-\tfrac{S_1}{S_0}\big) \subset \cA.
\]
This means that we may add to any acceptable position the payoff of a fully-leveraged portfolio --- a portfolio in which \textit{either} the purchase of the risky asset if fully financed by borrowing at the risk-free rate \textit{or} the purchase of the risk-free asset is fully financed by shorting the eligible asset --- without compromising acceptability. In particular, since $\cA$ is a cone, we must have
\[
\Span\big(1-\tfrac{S_1}{S_0}\big) \subset \cA.
\]
This shows that $\rho_{\cA,S}$ can be comonotonic only if every fully-leveraged position is acceptable. However, sound acceptance sets will tend to be sensitive to very risky fully-leveraged positions because such positions may imply a huge amount of downside risk. This suggests that comonotonicity cannot be expected to hold for risky eligible assets that deviate too much from the risk-free asset.

\medskip

When specialized to convex acceptance sets, the preceding theorem can be characterized by means of a simpler condition as follows. Recall that if $\cA$ is convex and $\rho_\cA$ is comonotonic, then $\cA$ is automatically a convex cone.

\begin{corollary}
\label{cor: characterization comonotonicity convex risk measures}
Assume that $\cA$ is convex and $\rho_\cA$ is comonotonic. Then, the following statements are equivalent:
\begin{enumerate}[(a)]
  \item $\rho_{\cA,S}$ is comonotonic.
  \item $\pm\big(1+\tfrac{\rho_{\cA,S}(1)}{S_0}S_1\big)\in\cA$.
  \item $S_1+\tfrac{S_0}{\rho_{\cA,S}(1)}\in\cA\cap(-\cA)$.
\end{enumerate}
\end{corollary}
\begin{proof}
Since $0\in\cA$ by conicity, it follows directly from Theorem~\ref{thm: characterization comonotonicity risk measures} that {\em (a)} implies {\em (b)}. To prove the converse implication, assume that {\em (b)} holds and take an arbitrary $X\in\cA$. Since, being convex and conic, $\cA$ is closed under addition, we infer from {\em (b)} that $X\pm(1+\tfrac{\rho_{\cA,S}(1)}{S_0}S_1)\in\cA$. That $\rho_{\cA,S}$ is comonotonic now follows from Theorem~\ref{thm: characterization comonotonicity risk measures}. The equivalence between {\em (b)} and {\em (c)} is a direct consequence of conicity.
\end{proof}

\medskip

In the setting of the above corollary, the set $\cA\cap(-\cA)$ is a linear space whose elements might be called ``risk invariants'' because adding them to a capital position $X$ has no impact on $\rho_\cA(X)$, i.e.
\[
\cA\cap(-\cA) = \{X\in L^\infty\mid \rho_\cA(X+Y)=\rho_\cA(Y), \ Y\in L^\infty \}.
\]
This is a direct consequence of the subadditivity of $\rho_\cA$. In this sense, the preceding result tells us that $\rho_{\cA,S}$ is comonotonic if and only if the payoff of the ``fully-leveraged'' position $S_1+\tfrac{S_0}{\rho_{\cA,S}(1)}$ is a risk invariant.

\medskip

As a consequence of the preceding corollary we see that, if the acceptance set $\cA$ satisfies the pointedness condition
\[
\cA\cap(-\cA)=\{0\},
\]
i.e.\ if there exists no nonzero risk invariant, then a risk measure $\rho_{\cA,S}$ cannot be comonotonic unless the eligible asset $S$ is risk-free. As illustrated by our examples in the next section, this situation is far from being exceptional. In fact, pointedness is the rule rather than the exception for the vast majority of acceptance sets used in applications.

\begin{corollary}
\label{cor: comonotonicity under pointedness}
Assume that $\cA$ is pointed. Then, $\rho_{\cA,S}$ is comonotonic only if
\[
S_1 = -\tfrac{S_0}{\rho_{\cA,S}(1)}.
\]
In particular, $\rho_{\cA,S}$ is never comonotonic if $S$ is a risky asset.
\end{corollary}

\subsubsection*{Characterizing comonotonicity under risk-free assets}

As already observed, it follows from Theorem~\ref{thm: characterization comonotonicity risk measures} that the comonotonicity of $\rho_{\cA,S}$ depends on the interplay between the acceptance set $\cA$ and the eligible asset $S$. Thus, in contrast to many other important properties of risk measures such as convexity or positive homogeneity, comonotonicity cannot be characterized by properties of the acceptance set alone. The aim of this brief section is to show that, if we restrict our attention to convex cash-additive risk measures, such a characterization is possible.

\begin{theorem}
\label{theo: comonotonicity in terms of A}
Assume that $\cA$ is convex. Then, the following statements are equivalent:
\begin{enumerate}[(a)]
  \item $\rho_\cA$ is comonotonic.
  \item For each $\cC\in\cC_{max}$ there is a positive linear functional $\pi_\cC:L^\infty\to\R$ such that $\pi_\cC\geq\pi_\cD$ on $\cC$ for all $\cD\in\cC_{max}$ and
\[
\cA = \bigcap_{\cC\in\cC_{max}}\{X\in L^\infty\mid \pi_\cC(X)\geq0\}.
\]
\end{enumerate}
In this case, for every $X\in L^\infty$ we have
\begin{equation}
\label{eq: dual representation comonotonic map}
\rho_\cA(X) = \max_{\cC\in\cC_{max}}\pi_\cC(-X).
\end{equation}
\end{theorem}
\begin{proof}
Assume that {\em (b)} holds and note that, for every $X\in L^\infty$, we must have
\[
\rho_\cA(X) = \inf\{m\in\R\mid \pi_\cC(X+m)\geq0, \ \cC\in\cC_{max}\} = \sup_{\cC\in\cC_{max}}\pi_\cC(-X).
\]
Since $\pi_\cC(X)\geq\pi_\cD(X)$ for all $\cD\in\cC_{max}$, we infer that $\rho_\cA=\pi_\cC$ on $\cC$. In conclusion, Proposition~\ref{prop: characterization comonotonicity decreasing maps} implies that $\rho_\cA$ is comonotonic.

\smallskip

Conversely, assume that $\rho_\cA$ is comonotonic. Since $\rho_\cA$ is convex, it follows from Proposition~\ref{prop: characterization comonotonicity decreasing maps} that for each $\cC\in\cC_{max}$ we find a positive linear functional $\pi_\cC:L^\infty\to\R$ such that $\rho_\cA=-\pi_\cC$ on $\cC$ and $\rho_\cA\geq-\pi_\cC$ on $L^\infty$. This immediately yields~\eqref{eq: dual representation comonotonic map}. The representation of $\cA$ in {\em (b)} is now a direct consequence of the fact that $\cA=\cA(\rho_\cA)$ by Proposition~\ref{prop: usual properties}.
\end{proof}

\smallskip

\begin{remark}
Theorem~\ref{theo: comonotonicity in terms of A} can be viewed as a generalization of the finite-dimensional characterization of comonotonicity in Rieger~(2017). This follows from Remark~\ref{rem: comonotonic sets and permutations}, where we observed that, in case of a finite probability space, one can represent maximal comonotonic sets by means of (finitely many) permutations.
\end{remark}


\section{Examples}
\label{three}

In this final section we illustrate our results by applying them to a variety of explicit acceptance sets that includes the most common examples. These concrete examples highlight that, for risky eligible assets, comonotonicity is the exception rather than the rule.


\subsubsection*{Capital adequacy based on Value-at-Risk}

The {\em Value-at-Risk} (VaR) of a capital position $X\in L^\infty$ at the level $\alpha\in(0,1)$ is defined by
\[
\VaR_\alpha(X) := \inf\{m\in\R\mid \probp(X+m<0)\leq\alpha\}.
\]
Note that $\VaR_\alpha(X)$ is, up to a sign, the upper $\alpha$-quantile of $X$. The corresponding acceptance set is the closed cone given by
\[
\accvar(\alpha) := \{X\in L^\infty\mid \VaR_\alpha(X)\leq0\} = \{X\in L^\infty\mid \probp(X<0)\leq\alpha\}.
\]
We are interested in studying the comonotonicity of the risk measure $S$-$\VaR_\alpha:L^\infty\to\R$ given by
\[
\mbox{$S$-$\VaR_\alpha(X)$} := \rho_{\accvar(\alpha),S}(X) = \inf\{m\in\R\mid \probp(X+\tfrac{m}{S_0}S_1<0)\leq\alpha\}.
\]

\smallskip

Since $\VaR_\alpha$ is well-known to be comonotonic, one easily sees that $S$-$\VaR_\alpha$ will be automatically comonotonic whenever $S$ is risk-free. In this case, $S$-$\VaR_\alpha$ will be, in fact, just a multiple of $\VaR_\alpha$. At the same time, it is not difficult to verify that the acceptance set $\accvar(\alpha)$ is not pointed in general and, therefore, Corollary~\ref{cor: comonotonicity under pointedness} does not apply to risk measures based on VaR-acceptability.

\medskip

The first result is derived by applying Theorem~\ref{thm: characterization comonotonicity risk measures} to VaR-acceptability and shows that $S$-$\VaR_\alpha$ can be comonotonic only if the payoff $S_1$ is constant with sufficiently high probability. Indeed, as values of $\alpha$ close to $0$ are the interesting ones from a practical perspective, the bound given in~\eqref{eq: necessary condition comonotonicity VaR} is close to $1$.

\begin{proposition}
\label{prop: necessary condition comonotonicity var}
Assume that $S$-$\VaR_\alpha$ is comonotonic. Then, we have
\begin{equation}
\label{eq: necessary condition comonotonicity VaR}
\probp\big(S_1=-\tfrac{1}{\VaR_\alpha(1/S_1)}\big) \geq 1-2\alpha.
\end{equation}
\end{proposition}
\begin{proof}
Since $0\in\accvar(\alpha)$, it follows from Theorem~\ref{thm: characterization comonotonicity risk measures} that $S$-$\VaR_\alpha$ can be comonotonic only if
\begin{equation}\label{eq:var1}
\probp\big(1+\tfrac{\mbox{\scriptsize $S$-$\VaR_\alpha(1)$}}{S_0}S_1<0\big)\leq\alpha \ \ \ \mbox{and} \ \ \ \probp\big(-1-\tfrac{\mbox{\scriptsize $S$-$\VaR_\alpha(1)$}}{S_0}S_1<0\big)\leq\alpha.
\end{equation}
By rearranging, the above inequalities are easily seen to imply
\[
\probp\big(S_1=-\tfrac{1}{\VaR_\alpha(1/S_1)}\big) =
1-\probp\big(S_1<-\tfrac{1}{\VaR_\alpha(1/S_1)}\big)-
\probp\big(S_1>-\tfrac{1}{\VaR_\alpha(1/S_1)}\big) \geq
1-2\alpha,
\]
where we used that $S$-$\VaR_\alpha(1)=S_0\VaR_\alpha(1/S_1)$.
\end{proof}

\smallskip

\begin{remark}
\label{ex: var not comonotonic}
Condition~\eqref{eq: necessary condition comonotonicity VaR} is generally not sufficient for comonotonicity. To see this, let $\{A,B,C\}$ be a measurable partition of $\Omega$ such that $\probp(A)=\probp(B)=\alpha$ and $\probp(C)=1-2\alpha$. Consider an eligible asset $S$ with $S_0>0$ and
\[
S_1=
\begin{cases}
S_0 & \mbox{on} \ A\cup C,\\
2S_0 & \mbox{on} \ B.
\end{cases}
\]
It is easy to verify that $\VaR_\alpha(1/S_1)=-1/S_0$ and, thus, \eqref{eq: necessary condition comonotonicity VaR} is satisfied. However, since
\[
-1_A\in\accvar(\alpha) \ \ \ \mbox{and} \ \ \ -1_A+\big(1-\tfrac{S_1}{S_0}\big) = -1_{A\cup B} \notin \accvar(\alpha),
\]
it follows from Theorem~\ref{thm: characterization comonotonicity risk measures} that $S$-$\VaR_\alpha$ is not comonotonic.
\end{remark}

\medskip

In view of the previous result, it is natural to wonder whether comonotonicity is at all compatible with risky eligible assets. The next proposition characterizes all the underlying probabilistic models where $S$-$\VaR_\alpha$ is comonotonic for some risky eligible asset.

\begin{lemma}
\label{prop:var comonotonic nonconstant S}
The following statements are equivalent:
\begin{enumerate}[(a)]
  \item There exists a risky eligible asset $S$ such that $S$-$\VaR_\alpha$ is comonotonic.
  \item There exists $A\in\cF$ such that $0<\probp(A)\leq\alpha$ and for every $B\in\cF$ we have
  \[
  \probp(B)\leq\alpha \ \implies \ \probp(A)+\probp(A^c\cap B)\leq\alpha.
  \]
\end{enumerate}
\end{lemma}
\begin{proof}
To prove that {\em (a)} implies {\em (b)}, assume that $S$-$\VaR_\alpha$ is comonotonic for some eligible asset $S$ with nonconstant payoff $S_1$. Consider the random variable
\[
Z = 1+\tfrac{\mbox{\scriptsize $S$-$\VaR_\alpha(1)$}}{S_0}S_1
\]
and note that, by~\eqref{eq:var1}, it satisfies
\[
\max\{\probp(Z<0),\probp(Z>0)\} \leq \alpha.
\]
Since $Z$ is nonconstant, one of the above probabilities must be strictly positive. Without loss of generality, assume that $\probp(Z<0)>0$ and set
\[
A = \{Z<0\} \in\cF.
\]
Now, take any $B\in\cF$ satisfying $\probp(B)\leq\alpha$. Since $-1_B\in\accvar(\alpha)$, Theorem~\ref{thm: characterization comonotonicity risk measures} implies that $Z-1_B$ must belong to $\accvar(\alpha)$ so that $\probp(Z<1_B)\leq\alpha$. Note that $\probp(Z<1)=1$ because $S$-$\VaR_\alpha(1)<0$. Then, it is easy to see that
\[
\probp(A)+\probp(A^c\cap B) = \probp(A\cup(A^c\cap B)) \leq \probp(Z<1_B) \leq \alpha.
\]
Hence, {\em (a)} implies {\em (b)}.

\smallskip

To prove the converse implication, assume that {\em (b)} holds and define the eligible asset $S$ by setting $S_0=1$ and
\[
S_1=
\begin{cases}
1 & \mbox{on} \ A^c,\\
2 & \mbox{on} \ A.
\end{cases}
\]
Moreover, consider the random variable
\[
Z = 1+\tfrac{\mbox{\scriptsize $S$-$\VaR_\alpha(1)$}}{S_0}S_1.
\]
It is easy to verify that $S$-$\VaR_\alpha(1)=-1$ so that $Z=-1_A$. Take now an arbitrary $X\in\accvar(\alpha)$. By applying {\em (b)} to $B=\{X<0\}$ we obtain
\[
\probp(X+Z<0) = \probp(A\cap\{X<1_A\})+\probp(A^c\cap\{X<1_A\}) \leq \probp(A)+\probp(A^c\cap\{X<0\}) \leq \alpha.
\]
In addition, we easily see that
\[
\probp(X-Z<0) = \probp(X<-1_A) \leq \probp(X<0) \leq \alpha.
\]
As a result, Theorem~\ref{thm: characterization comonotonicity risk measures} implies that $S$-$\VaR_\alpha$ is comonotonic and we conclude that {\em (b)} implies {\em (a)}. This completes the proof of the equivalence.
\end{proof}

\medskip

The preceding result has the following unambiguous consequence when specified to the common setting of a {\em nonatomic} probability space, i.e.\ a probability space that supports random variables with any prescribed distribution: Risk measures based on $\VaR$-acceptability are never comonotonic unless the eligible asset is risk-free.

\begin{proposition}
\label{cor: var risk measures nonatomic}
Assume that $(\Omega,\cF,\probp)$ is nonatomic. Then, $S$-$\VaR_\alpha$ is comonotonic if and only if $S$ is risk-free.
\end{proposition}
\begin{proof}
The ``if'' implication is always true. To prove the ``only if'' implication, it suffices to note that condition {\em (b)} in Lemma~\ref{prop:var comonotonic nonconstant S} is never satisfied in a nonatomic setting. Indeed, if $\probp(A)\leq \alpha$, then we can always find $B\in\cF$ such that $B\subset A^c$ and $\alpha-\probp(A)<\probp(B)<\alpha$ by nonatomicity. Since we easily have
\[
\probp(A)+\probp(A^c\cap B) = \probp(A)+\probp(B) > \alpha,
\]
Lemma~\ref{prop:var comonotonic nonconstant S} tells us that $S$-$\VaR_\alpha$ can be comonotonic only if $S_1$ is constant.
\end{proof}

\smallskip

\begin{remark}
It is well-known that $\VaR_\alpha$ fails to be subadditive. However, being comonotonic, it satisfies
\[
\VaR_\alpha(X+Y) = \VaR_\alpha(X)+\VaR_\alpha(Y)
\]
for any comonotone $X,Y\in L^\infty$. This allows to control the capital required for an aggregated position of comonotone random variables by means of the individual capital requirements. Since $S$-$\VaR_\alpha$ is in general not comonotone if $S$ is a risky asset, one may wonder whether the capital required for an aggregated position of comonotone random variables can still be controlled in terms of the individual capital requirements or not. Here, we show that the undesirable situation
\[
\mbox{$S$-$\VaR_\alpha(X+Y)$} > \mbox{$S$-$\VaR_\alpha(X)$}+\mbox{$S$-$\VaR_\alpha(Y)$}
\]
is possible also for comonotone $X,Y\in L^\infty$, so that summing up the individual capital requirements of comonotone random variables does not help find a bound for the capital required for the aggregated position.

\medskip

We provide an example in the setting of Example~\ref{ex: var not comonotonic} above. If we consider the comonotone random variables
\[
X =
\begin{cases}
-2 & \mbox{on $A$}\\
-3 & \mbox{on $B$}\\
2 & \mbox{on $C$}
\end{cases}
 \ \ \ \mbox{and} \ \ \ Y =
\begin{cases}
-4 & \mbox{on $A$}\\
-9 & \mbox{on $B$}\\
0 & \mbox{on $C$}\\
\end{cases},
\]
then it is not difficult to show that
\[
\mbox{$S$-$\VaR_\alpha(X+Y)$} = 6 > \tfrac{3}{2}+4 = \mbox{$S$-$\VaR_\alpha(X)$}+\mbox{$S$-$\VaR_\alpha(Y)$}.
\]
\end{remark}


\subsubsection*{Capital adequacy based on Expected Shortfall}

The {\em Expected Shortfall} (ES) of a capital position $X\in L^\infty$ at the level $\alpha\in(0,1)$ is defined by
\[
\ES_\alpha(X) := \frac{1}{\alpha}\int_0^\alpha\VaR_\beta(X)\,d\beta.
\]
The corresponding acceptance set is the closed convex cone defined by
\[
\acces(\alpha) := \{X\in L^\infty\mid \ES_\alpha(X)\leq0\}.
\]
We aim to characterize comonotonicity for the risk measure $S$-$\ES_\alpha:L^\infty\to\R$ given by
\[
\mbox{$S$-$\ES_\alpha(X)$} := \rho_{\acces(\alpha),S}(X) = \inf\{m\in\R\mid \ES_\alpha(X+\tfrac{m}{S_0}S_1)\leq0\}.
\]

\medskip

We show that risk measures based on $\ES$-acceptability are comonotonic if, and only if, the eligible asset is risk-free. This result will be a direct consequence of the following lemma.

\begin{lemma}
\label{lem:avar}
For every nonconstant $X\in L^\infty$ we have $\ES_\alpha(X)>-\E[X]$.
\end{lemma}
\begin{proof}
Since $\VaR_\beta(X)$ is, for every $X\in L^\infty$, a decreasing function of $\beta$, it follows that
\[
\ES_\alpha(X)-\int_0^\alpha\VaR_\beta(X)\,d\beta = \frac{1-\alpha}{\alpha}\int_0^\alpha\VaR_\beta(X)\,d\beta \ge
(1-\alpha)\VaR_\alpha(X) \ge
\int_\alpha^1\VaR_\beta(X)\,d\beta,
\]
with equality between the left- and the right-hand side if and only if $\VaR_\beta(X)=\VaR_\alpha(X)$ for all $\beta\in(0,1)$ or, equivalently, if and only if $X$ is constant. Hence, by rearranging, we obtain
\[
\ES_\alpha(X) \geq \int_0^\alpha\VaR_\beta(X)\,d\beta+\int_\alpha^1\VaR_\beta(X)\,d\beta = \int_0^1\VaR_\beta(X)\,d\beta = -\E[X],
\]
with equality if and only if $X$ is constant.
\end{proof}

\medskip

The preceding lemma implies that acceptance sets based on ES are pointed and, thus, we are in a position to apply Corollary~\ref{cor: comonotonicity under pointedness} and conclude that risk measures based on ES-acceptability fail to be comonotonic unless the eligible asset is risk-free.

\begin{proposition}
\label{prop: ES risk measures}
The risk measure $S$-$\ES_\alpha$ is comonotonic if and only if $S$ is risk-free.
\end{proposition}
\begin{proof}
The ``if'' implication is clear since $S$-$\ES_\alpha$ is simply a multiple of $\ES_\alpha$ in this case and $\ES_\alpha$ is well known to be comonotonic. To prove the ``only if'' implication, note that
\begin{equation}
\label{eq: ES risk measures}
\ES_\alpha(X)+\ES_\alpha(-X) > -\E[X]+\E[X] = 0
\end{equation}
holds for every nonconstant $X\in L^\infty$ by Lemma~\ref{lem:avar}. As a result, it follows that the acceptance set $\acces(\alpha)$ satisfies the pointedness condition
\[
\acces(\alpha)\cap(-\acces(\alpha))=\{0\}.
\]
To see this, assume that $X\in\acces(\alpha)\cap(-\acces(\alpha))$ so that $\ES_\alpha(X)\leq0$ and $\ES_\alpha(-X)\leq0$ both hold. Since $\ES_\alpha(X)+\ES_\alpha(-X)\leq0$, it follows from~\eqref{eq: ES risk measures} that $X$ must be constant. However, the only constant random variable belonging to $\acces(\alpha)\cap(-\acces(\alpha))$ is clearly the zero random variable. In conclusion, $\acces(\alpha)$ is pointed and Corollary~\ref{cor: comonotonicity under pointedness} implies that $S$-$\ES_\alpha$ is comonotonic only if $S$ is risk-free.
\end{proof}


\subsubsection*{Capital adequacy based on distortion risk measures}

We denote by $\cP([0,1])$ the set of all probability measures $\mu:[0,1]\to[0,1]$. The {\em distortion risk measure} associated to $\mu\in\cP([0,1])$ is the map $\CD_\mu:L^\infty\to\R$ defined by
\[
\CD_\mu(X) := \int_0^1\ES_\alpha(X)\,\mu(d\alpha).
\]

\smallskip

Here, as is commonly done, we extend ES by setting
\[
\ES_0(X):=-\inf\{m\in\R\mid X\geq m\} \ \ \ \mbox{and} \ \ \ \ES_1(X):=-\E[X].
\]
The corresponding acceptance set is the closed convex cone given by
\[
\cA_{\CD}(\mu) := \{X\in L^\infty\mid \CD_\mu(X)\leq0\}.
\]

\medskip

The class of acceptance sets based on distortion risk measures is huge and includes, in a nonatomic setting, all the acceptance sets that are convex, law-invariant and compatible with comonotonicity. This is made precise in the next proposition, which is a direct consequence of Theorem~4.93 in F\"{o}llmer and Schied~(2011).

\begin{proposition}
Assume that $(\Omega,\cF,\probp)$ is nonatomic and $\cA$ is convex and law-invariant. Then, the following statements are equivalent:
\begin{enumerate}[(a)]
  \item $\rho_\cA$ is comonotonic.
  \item $\cA=\cA_{\CD}(\mu)$ for some $\mu\in\cP([0,1])$.
\end{enumerate}
\end{proposition}

\smallskip

\begin{remark}
In a nonatomic setting, distortion risk measures can be equivalently identified, up to a sign, with Choquet integrals associated with concave distortion functions, see Theorem~4.70 in F\"{o}llmer and Schied~(2011).
\end{remark}

\medskip

We aim to characterize comonotonicity for the risk measure $S$-$\CD_\mu:L^\infty\to\R$ given by
\[
\mbox{$S$-$\CD_\mu(X)$} := \rho_{\cA_{\CD}(\mu),S}(X) = \inf\{m\in\R\mid \CD_\mu(X+\tfrac{m}{S_0}S_1)\leq0\}.
\]

\medskip

The next proposition shows that distortion-based risk measures are never comonotonic unless they reduce to standard expectations or the eligible asset is taken to be risk-free.

\begin{proposition}
\label{prop: distortions}
The risk measure $S$-$\CD_\mu$ is comonotonic if and only if one of the following conditions holds:
\begin{enumerate}[(i)]
  \item $\mu(\{1\})=1$ (so that $\CD_\mu(X)=-\E[X]$ for all $X\in L^\infty$).
  \item $S$ is risk-free.
\end{enumerate}
\end{proposition}
\begin{proof}
The ``if'' implication is clear since under any of the two conditions the risk measure $S$-$\CD_\mu$ is simply a multiple of $\CD_\mu$, which is comonotonic. In particular, if $\mu({1})=1$, then we clearly have $\CD_\mu(X)=-\E[X]$ for any $X\in L^\infty$ so that
\[
\mbox{$S$-$\CD_\mu(X)$} = \inf\{m\in\R\mid \E[X+\tfrac{m}{S_0}S_1]\geq0\} = -\tfrac{S_0}{\E[S_1]}\E[X] = \tfrac{S_0}{\E[S_1]}\CD_\mu(X)
\]
for all $X\in L^\infty$. To prove the ``only if'' implication, assume that $S$-$\CD_\mu$ is comonotonic but $\mu(\{1\})<1$. Then, it follows from Lemma~\ref{lem:avar} that any nonconstant $X\in L^\infty$ satisfies $\CD_\mu(X)>-\E[X]$ and therefore
\[
\CD_\mu(X)+\CD_\mu(-X) > -\E[X]+\E[X] = 0.
\]
As in the proof of Proposition~\ref{prop: ES risk measures}, this yields the pointedness condition
\[
\cA_{\CD}(\mu)\cap(-\cA_{\CD}(\mu)) = \{0\}.
\]
Hence, we can apply Corollary~\ref{cor: comonotonicity under pointedness} and conclude that $S$ must be risk-free. This completes the proof of the ``only if'' implication.
\end{proof}


\section{Conclusions}
\label{four}

Our discussion and results allow the following two conclusions:
\begin{itemize}
  \item {\bf Comonotonicity heavily depends on the eligible asset}. In contrast to other important properties of risk measures such as convexity, subadditivity, and positive homogeneity, our results show that comonotonicity cannot be characterized in terms of properties of the underlying acceptance set alone but critically depends on the particular choice of the eligible asset. Recall that acceptability is the key ``regulatory'' objective and that the eligible asset is simply a vehicle to achieve it. Hence, it would seem that, within the context of capital adequacy, comonotonicity is an incidental, rather than a fundamental, property.
  \item {\bf Comonotonicity is typically incompatible with risky eligible assets}. Comonotonicity is compatible with only a limited range of eligible assets, which typically need to be close to risk-free. In fact, for many important acceptance sets --- such as those based on Value-at-Risk or distortion risk measures like Expected Shortfall in nonatomic probability spaces --- comonotonicity is compatible only with a risk-free eligible asset. On the one side, this means that imposing a comonotonicity requirement on a regulatory risk measure implicitly restricts the choice of vehicles to reach acceptability. This runs contrary to the premise that acceptability is the primary regulatory concern. In fact, restricting the choice of eligible assets ultimately reduces capital efficiency as pointed out in Artzner et al.~(2009). On the other side, the incompatibility of comonotonicity with risky eligible assets means that no operationally relevant comonotonic risk measure can exist in an economy without a risk-free asset.
\end{itemize}

These findings qualify and arguably call for a critical appraisal of the meaning and the role of comonotonicity within a capital adequacy context.


\end{document}